\documentclass[traditabstract]{aa}

\usepackage{txfonts}
\usepackage{natbib}
\usepackage{epsfig,times,lscape,rotating}
\usepackage{color}
\usepackage{amssymb}

\begin{document}
\title{A high-order Godunov scheme for global 3D MHD accretion disks simulations. I.
The linear growth regime of the magneto-rotational instability}
\author{M. Flock$^1$, N. Dzyurkevich$^1$, H. Klahr$^1$, A. Mignone$^2$}
\offprints{flock@mpia.de}
\institute{$^1$Max-Plank Institute for Astronomy,
              K\"onigstuhl 17, 69117 Heidelberg,
              Germany\\
                    $^2$Dipartimento di Fisica Generale, Universit\'{a} degli Studi di Torino, via Pietro Giuria 1, 10125 Torino, Italy
}
\date{\today}
\authorrunning{Flock et al.}
%
%\titlerunning{\ Inter code comparison }
%
\abstract{
We employ the PLUTO code for computational astrophysics to
assess and compare the validity of different numerical algorithms
on simulations of the magneto-rotational instability in 3D accretion disks.
In particular we stress on the importance of using a consistent upwind
reconstruction of the electro-motive force (EMF) when using the
constrained transport (CT) method to avoid the onset of numerical
instabilities.
We show that the electro-motive force (EMF) reconstruction in the classical 
constrained transport (CT) method for Godunov schemes drives a numerical instability. 
The well-studied linear growth of magneto-rotational instability (MRI) is used as a benchmark for an inter-code
comparison of PLUTO and ZeusMP.
We reproduce the analytical results for linear MRI growth in 3D global MHD simulations and present a robust and accurate Godunov code which can be used for 3D accretion disk simulations in curvilinear coordinate systems.
}
\maketitle
%------------------------------------------------------------------
%  Definition of fonts
\def\kpc{{\ \rm kpc}}
\def\kms{{\ \rm km\ s^{-1}}}
\def\ri{r_{\rm i}}
\def\ro{r_{\rm out}}
\def\rs{r_{\star}}
\def\mj{\dot{M}_{\rm jet}}
\def\ms{{M}_{\star}}
\def\vk{v_{\rm K}}
\def\rj{R_{\rm jet}}
\def\rk{R_{\kappa}}
\def\bp{B_{\rm P}}
\def\bh{B_{\phi}}
\def\jp{\vec{B}_{\rm P}}
\def\cm{{\rm cm}}
\def\msun{{\rm M}_{\sun}}
\def\rsun{{\rm R}_{\sun}}
\renewcommand{\vec}[1]{\mbox{\boldmath$#1$}}
\def\gsim{\lower.4ex\hbox{$\;\buildrel >\over{\scriptstyle\sim}\;$}} 
\def\lsim{\lower.4ex\hbox{$\;\buildrel <\over{\scriptstyle\sim}\;$}}
\def\ab{\langle{\vec{B}\rangle}}
\def\au{\langle{\vec{u}\rangle}}
\def\A{Alfv\'en}
%-----------------------------------------------------------------------------
%-----------------------------------------------------------------------------
%%%%%%%%%%%%%%%%%%%%%%%%%%%%%%%%%%%%%%%%%%%%%%%%%%%%%%%%%%%%%%%%
\section{Introduction}
%%%%%%%%%%%%%%%%%%%%%%%%%%%%%%%%%%%%%%%%%%%%%%%%%%%%%%%%%%%%%%%%
Most classical work on the MRI has been conducted using local shearing sheet 
simulations \citep{bra95,haw95,san04}. Nevertheless, recent development has shown that critical
questions about turbulence in accretion disks can only be addressed in global
simulations, e.g. the saturation level of turbulence, the global evolution
and the role of the global field configuration. 
So far most existing codes which can handle global disk simulations are
based on a Zeus like finite difference scheme \citep{haw00,fro06,fro09}.
Thus, it would be very beneficial to have an independent method to attack the
open questions in Magneto-Hydrodynamic disk simulations.
The recently introduced MHD Godunov code PLUTO \citep{mig07} brought new insight into the
field of jet propagation and for instance MHD simulations in the solar
system, e.g. the space weather prediction.
The latter example has the benefit that the predictions obtained with the
numerical method can be tested against observation. 
A number of investigators have recognized the importance of using
conservative Godunov-type upwind schemes rather than non-conservative finite
difference algorithms which do not make use of the characteristic of the
associated Riemann Problem.
Clearly accretion disk calculations using conservative schemes are
of great interest.
As we want to apply our simulations to study the physical conditions for the
planet formation process in circumstellar disks, we need not only the
dynamical evolution of the disk, but also the detailed thermal and chemical
conditions of the disk gas. 
None of the global disk simulations so far have been performed with tracing the
temperature changes in the gaseous disk. 
The first result in this direction is 
comparing MRI linear growth rates including radiation transport which was 
achieved in local-box simulations \citep{fla09}.
It is widely accepted that the magneto-rotational instability (MRI) drives the turbulence in proto-planetary disks,
leading to gas accretion onto the star as well as radial and vertical
mixing. The efficiency of MRI-driven turbulence depends on the ionization
degree of the gas. 
To determine the latter, thermal evolution including
heat input from magnetic field dissipation has to be
taken into account.
Therefore we need an accurate conservative numerical scheme for treating such MHD
processes as magnetic diffusion and reconnection in the presence of the small-scale
turbulence.
One of the most important challenges is an accurate and effective way to
control and monitor the constraint of the divergence-free magnetic field.
In the last years, several approaches on the numerical treatment of ${\rm div}(\vec{B})$
have been proposed. 
Two main classes for Godunov schemes have been established over the
years. First, Powell's
"Eight Waves" method uses the modified Euler equations with specific source terms and
allows for magnetic monopoles treated as the $8^{\rm{th}}$ wave (Powell 1994).
This idea is similar to the elliptical cleaning
method \citep{ded02} which damps numerical monopoles. 
The second class is the Constrained Transport method, initially used as MocCT in ZEUS and 
then adapted to Godunov schemes \citep{bal99}. Here, the magnetic
field components are always reprojected on the staggered grid and ${\rm div}(\vec{B}) =0$ is kept
within machine accuracy. 
The classical CT method for Godunov schemes \citep{bal99} uses an arithmetic averaging of the
electro-motive forces (EMF).
This average proves to be not consistent with the plane parallel propagation in upwind schemes \citep{lon00}.
 \citep{lon04} have proposed a new way, named {\it upwind
  constrained transport} (UCT), of combining CT and a consistent upwind EMF reconstruction.
\citet{gar05} also developed a consistent method for the CT implementation in the ATHENA code.
The PLUTO code offers a choice of the methods described above. 
The PLUTO code \citep{mig07} is a conservative multi-dimensional and multi-geometry code, with applications for Newtonian, relativistic, MHD and relativistic MHD fluids. 
The possibility of switching between several
numerical modules in PLUTO allows us to perform a broad variety of test problems
and choose the module with the best convergence and stability. 
E.g. the code offers the possibility of using
Riemann solvers of ROE, HLLD, HLLC, HLL and Lax-Friedrich Rusanov.
In this paper we investigate the applicability of the MHD methods for 3D curvilinear global
simulations of the turbulent disk.
To have comparable and reliable results, simulations with the ZEUS code are
included in this paper, representing the group of the finite difference codes.
ZeusMP version 1b has been used successfully for 3D MHD global
 simulations of astrophysical disks. ZeusMP \citep{sto92,hay06} has a classical staggered
discretization of magnetic and electric vector fields (MoC CT).  
An identical model is set up in the actual ZeusMP code as well as the PLUTO
3.
We examine the performance of the different numerical schemes by comparing the growth 
of the MRI with the expectations from linear analysis.
The PLUTO code has successfully reproduced a series of standard MHD tests, including
the rotating shock-tube, the fast rotor and the blast wave solution \citep{mig07}.
In addition, a magnetized accretion torus in
2.5 dimensions is demonstrated.\\ 
We are testing the PLUTO code specifically for the linear MRI.
Our 3D disk setup allows to reproduce exactly the
conditions for MRI typical for accretion disk studies.
At the same time, unphysical results can
easily be discovered by analytical linear MRI analysis.
In section 2 we describe the differences in the numerical schemes of ZEUS and PLUTO. Section 3
presents our disk model used for linear MRI simulation and the measurement
method. In Section 4 we show the results of the inter-code comparison.
A convergence check for the chosen code configuration is investigated in section
5. Conclusion and outlook are given in section 6. 
\section{Numerics}
\subsection {MHD Equations}
The equations of ideal magnetohydrodynamics written for Godunov schemes in conservative
form are:
 \begin{equation}
 \frac{\partial \rho\vec{u} }{\partial t} + \nabla \cdot (\rho\vec{u}\vec{u} -\vec{B}\vec{B})= - \nabla P^{*} +\rho  \nabla \Psi,
\label{1}
\end{equation}
\begin{equation}
\frac{\partial\rho}{\partial t} +  \nabla \cdot (\rho \vec{u}) = 0,
\label{2}
\end{equation}
\begin{equation}
\frac{\partial \vec{B}}{\partial t} + \nabla\cdot(
\vec{u}\vec{B} -  \vec{B}\vec{u})=0,                 
\label{3}
\end{equation}
and
\begin{equation}
 \frac{\partial E }{\partial t} + \nabla \cdot (( E+ P^{*})\vec{u} - \vec{B} (\vec{B}\vec{u}))= \rho \vec{u} \nabla \Psi
\label{4}
\end{equation}
where $\rho$ is the gas density, $\rho \vec{u}$ is the momentum density,
$\vec{B}$ is the magnetic field, and $E$ is the total energy density.
The total pressure is a sum of magnetic and gas pressure, $P^{*}=P+(
\vec{B}\cdot \vec{B})/2$.
Total energy density is connected to internal energy $\epsilon$ as $E=\epsilon + \rho
u^2/2 + B^2/2$. Gravitational potential $\Psi=1/R$ is chosen in the
normalized form to describe the sheared azimuthal flow of the gas, reproducing the Kepler rotation.
We use cylindric coordinates in our models with the notation: $(R,\phi, Z)$.
The equations are solved using  uniform grids.
In our runs we take an adiabatic
approximation for the Equation of State, $P=(\gamma-1)\epsilon$, with
$\gamma=5/3$.\\
Treatment in ZEUS: The equations (1-4) are solved by ZeusMP in a non-conservative way, i.e. the
momentum conservation will be treated in the following way,
\begin{equation}
\rho \frac{{\rm d} \vec{u} }{{\rm d} t}= - \nabla P +\rho  \nabla \Psi + \frac{1}{4\pi}[\nabla\times{}\vec{B}]\times{}\vec{B}.
\label{1a}
\end{equation}
As the vector components of velocity and magnetic fields are stored at the grid interfaces,
fluxes can be calculate directly in ZEUS. 
Density and pressure are stored at the cell center and 
the vector values are at the interfaces.
The scheme is second order in space and time. For time integration, ZEUS
uses a Hancock scheme.
Contrary to Zeus-type finite difference schemes, a Godunov type scheme, such as
represented by the PLUTO code, follows a conservative model. 
Therefore the temperature can be exactly calculated from the total energy
conservation. 
In general all quantities are stored at
the cell center and an 2nd order Runge-Kutta (RK) iterator is employed. Prior to flux
calculation, variables are reconstructed at the grid interface, which
implies interpolation with a limited slope.
The resulting two different states at the interfaces allow to solve the
Riemann problem, which is the main feature of a Godunov code.
CFL value is 0.25.

\subsection{Magnetic fields in Godunov schemes}
In case of MHD, cell-centered Godunov schemes have in general problems with occurring numerical magnetic monopoles \citep{bal99}.
Powell's "8 wave" method includes this numerical magnetic monopoles as source terms.
In the constrained transport (CT) MHD method, the magnetic
fields are located at the interfaces. % and interpolated at the cell-centers. 
Usually, this method allows to handle the magnetic monopoles at machine accuracy. 
After updating the staggered magnetic field, the 
cell centered variables will be interpolated from the new interface values.
The mainly used configuration in curvilinear coordinates systems consists of the second order upwind scheme
with piece-wise linear reconstruction and RK 2 for time evolution. 
In our inter-code comparison we test Powell's "8 Waves" method, the CT scheme and the 
different MHD Riemann solvers of Roe \citep{car97}, HLLD \citep{miy05}, HLL and Lax-Friedrich Rusanov.
We must note that
for the most frequently used configurations the ZEUS code has a speed up of factor 4. 
This factor appears because of the additional time step in the Runge Kutta time integration and the solving of the Riemann problem.

\begin{figure}
%\hspace{-0.6cm}
\begin{minipage}{5cm}
\psfig{figure=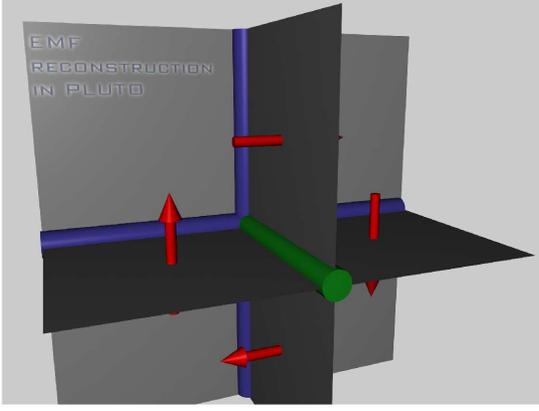,scale=0.50}
\end{minipage}
\label{act}
\caption{The reconstruction of the EMF (green axis) from four Godunov fluxes
  (red arrows) in case of arithmetic average of fluxes (ACT)  }
\end{figure}

\subsection{CT EMF reconstruction}
The solenoidality  of magnetic fields is difficult to represent correctly in Godunov schemes.
The simple arithmetic average to reconstruct the EMF is presented in Fig. 1 and
follows the Eq.(7)-Eq.(9) in \citet{bal99}. 
We share the opinion presented in \citet{lon00,lon04}, 
that such a simple hybrid method as ACT in Godunov scheme is not solving the monopole problem. 
The arithmetical CT algorithm is not consistent with the integration algorithm for plane-parallel,
grid-aligned flows, as in our case with the dominating Kepler rotation.
In PLUTO, the upwind CT scheme (UCT) follows \citet{gar05} and is based on the work of \citet{lon00,lon04}.
The main difference with respect to ACT is to construct the EMF update in such away that it is recovering the proper
directional biasing \citep{gar05}. The first method is here
called UCT. 
$$\hat{\varepsilon}_{z,i+1/2,j+1/2}=\frac{1}{2}(\varepsilon_{z,i+1/2,j}+\varepsilon_{z,i+1/2,j+1}+\varepsilon_{z,i,j+1/2}+\varepsilon_{z,i+1,j+1/2})$$
$$-\frac{1}{4}(\varepsilon_{z,i,j}+\varepsilon_{z,i,j+1}+\varepsilon_{z,i+1,j}+\varepsilon_{z,i+1,j+1})$$
The term $\varepsilon_{z,i+1/2,j+1/2}$ is taken from the Godunov fluxes as illustrated
in Fig. 1. $\varepsilon_{z,i,j}$ are the finite volume electric fields
calculated from the cell center values.
In addition to this method, we also test the EMF corner-integration, here
called $\rm UCT_{CONTACT}$:
$$\epsilon_{z,i+1/2,j+1/2}=\frac{1}{4}(\epsilon_{z,i+1/2,j}+\epsilon_{z,i+1/2,j+1}+\epsilon_{z,i,j+1/2}+\epsilon_{z,i+1,j+1/2})
$$ $$+\frac{\delta y}{8}( (\frac{\partial \epsilon_{z}}{\partial
y})_{i+1/2,j+1/4}-(\frac{\partial \epsilon_{z}}{\partial y})_{i+1/2,j+3/4})
$$ $$+\frac{\delta x}{8}( (\frac{\partial \epsilon_{z}}{\partial
x})_{i+1/4,j+1/2}-(\frac{\partial \epsilon_{z}}{\partial x})_{i+3/4,j+1/2})
$$
with
$$(\frac{\partial \epsilon_{z}}{\partial x})_{i+1/4,j}= \frac{2}{\delta
x}(\epsilon_{z,i+1/2,j} - \epsilon_{z,i,j})$$
 and 
$$\left(\frac{\partial \epsilon_{z}}{\partial y}\right)_{i+1/2,j+1/4}= \left\{
\begin{array}{ll} 
(\partial \epsilon_z / \partial y)_{i,j+1/4} &  {\rm for}\  v_{x,i+1/2,j} > 0,\\
(\partial \epsilon_z / \partial y)_{i+1,j+1/4} & {\rm for}\  v_{x,i+1/2,j} < 0,\\
\frac{1}{2} ( (\partial \epsilon_z / \partial y)_{i,j+1/4} \\
+(\partial \epsilon_z / \partial y)_{i+1,j+1/4} ), &  \rm{otherwise}.\\
\end{array} \right)   $$
The latter  shows in general the same MRI evolution
and therefor we concentrate on testing of UCT. 
\section{Linear MRI in global disk}
The magneto-rotational instability
(MRI) \citep{bal91,haw91,bal98}
is believed to occur in astrophysical disks. 
The analytical description of its linear stage has been discovered by \citet{bal91}.% and is valid in local approximations of shearing flows.    
Local shearing box simulations have confirmed the analytical expectation of
the linear MRI growth \citep{bra95,haw95}. Global simulations and also
nonlinear evolution of the MRI was presented in \citet{haw91}. 
Our simplified setup here is representative for a global proto-planetary disk 
and made for a modest radial domain.
\subsection{Global setup}
Our disk model is not representing the whole 3D accretion disk, 
but a part which is large enough to fulfill global properties. 
The gas is rotating differentially with the Keplerian shear,
$u_{\phi}^2=\frac{1}{R}$.
There is
no vertical stratification in both rotation and gas density.      
Initial temperature and density are constant in the entire disk patch,
$\rho=\rho_{0}$,  $T=T_{0}$, $c_0=0.1$. Therefore the orbital speed at the inner boundary
in units of the sound speed is 10.
The domain extends pi/3 radians in the azimuthal direction, $\rm{Z}=\pm0.5$ and from 1 to 4 units in the radial
direction.  An annulus of uniform vertical magnetic field is placed
at radii between 2 and 3 units, avoiding the radial boundaries so
that any effects of the boundary conditions are reduced.
The resolution is $[R,\phi,Z]=[128,64,64]$.
For both codes we use identical random generated velocity
perturbations ($10^{-4}u_{\rm Kep}$) for radial and vertical
velocities as the initial seed for the turbulence.
Boundary conditions are periodic for the vertical and azimuthal direction, and
zero gradient for the radial one.
Using the analytical prediction for the critical MRI wavelength, 
we choose the uniform vertical magnetic field strength in order to obtain the number of
fastest-growing wavelengths fitting in the domain heigh, $B_{z}=0.05513/n$ where
$n=1,2,...8$ (after equation 2.32 in \citet{bal91}).
The critical wavelength and the growth rate can be directly
measured from our simulations and compared with the analytical prediction.
% RESULT ##############

\subsection{Limits of the MRI growth rate}
Before using the growth rate as numerical benchmark, 
the physical processes limiting the growth rate in the linear MRI 
have to be considered. When magneto-rotational instability sets in, the
magnetic fields are amplified with $B=B_{0}\exp({\gamma t})$ until they reach
the saturation. Here we use a standard notation where $\gamma$ is the MRI growth rate.
As it is mentioned in \citet{haw91}, the absolute limit of the growth rate
for ideal MHD is given in case of zero radial wave vectors $q_r = 0$ with the
normalized wave vector $q_z$,
\begin{equation}
q_z=k_z\sqrt{16/15}\nu_A/\Omega.
\end{equation} 
with the Alv\'en speed $\nu_A=B/\sqrt{4\pi\rho}$.
Then, the critical mode $q_z=0.97$ grows
with $\gamma = 0.75 \Omega$, visible in Fig. 2.
Like in Balbus \& Hawley (1991b) we do not a priori know the
 radial wavenumber in our simulations. In Fig. 2 we plotted the
growth rate for the amplitude of different vertical modes.
The simulation is made for $B_{z}=0.05513/4$ with HLLD solver for 
resolution 128x64x64 (see also Table 2, $\rm{HLLD}_{\rm{LIN}}$).
We have obtained the MRI growth rates which are comparable with \citet{haw91},(Fig. 8).
The addition limiters of the MRI growth rate are viscous and resistive
dissipation \citep{pes08}.
We interpret the differences between the results, such as number of resolved MRI modes and
the growth rate, as the effect of numerical dissipation which is individual for each solver. 
\begin{figure}
\hspace{-0.6cm}
\begin{minipage}{5cm}
\psfig{figure=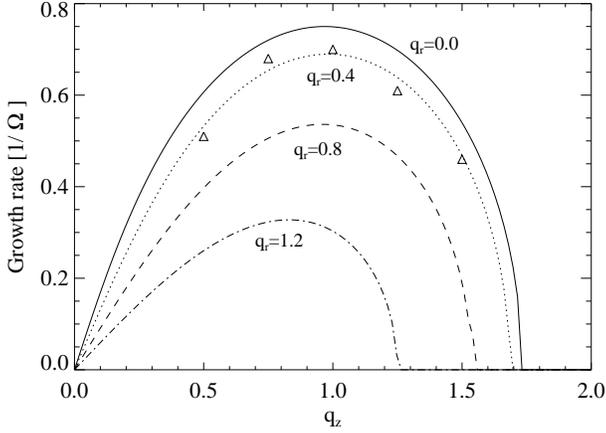,scale=0.50}
\end{minipage}
\label{growth}
\caption{Analytical growth rate is plotted for different $q_r$ (solid to
  dash-doted lines). The
triangles present the measured MRI growth rates from our simulation with
initial axisymmetric white noise disturbance from Model $\rm{HLLD}_{LIN}$ 128x64x64
with n=4. The values are measured for each radii and averaged. The results are similar to those
in \citet{haw91},(Fig. 8) }
\end{figure}
\begin{figure}
\hspace{-0.6cm}
\begin{minipage}{5cm}
\psfig{figure=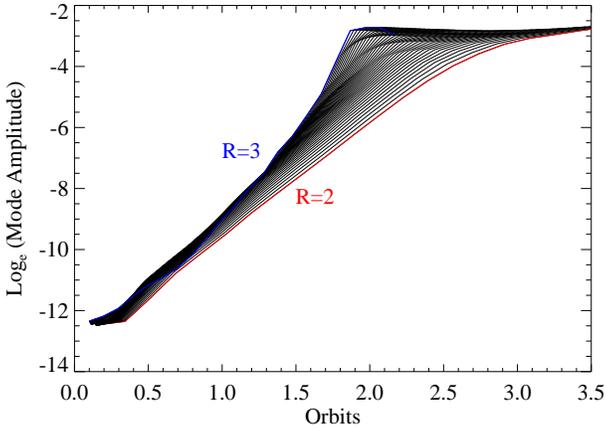,scale=0.50}
\end{minipage}
\label{growthall}
\caption{Maximal amplitude growth of the critical MRI mode is plotted for
  the 42 radial slices between radii $R=2$ and $R=3$ (solid lines). The setup is constructed to excite the mode
  with $\lambda_{\rm crit}=H$ ($n=1$). 
For the radii in the middle of MRI-active
  domain,
  the MRI growth rate reaches the value about $\gamma = 0.75 \Omega$. At the outer edge of the
  radial domain, the growth rate becomes higher than $ 0.75 \Omega$. We
  suggest that the MRI modes there
  are affected from the inner part where MRI already breaks into nonlinear evolution.
}
\end{figure}
In case of MRI, the numerical dissipation is the reason why we need some
minimal amount of the grid
cells to resolve the fastest growing MRI mode. \citet{haw95} claim to need at
least 5 grid cells per fastest growing mode. Generally, it holds that the more
grid cells we use per critical mode the less is the numerical dissipation.  
But beside increasing resolution, changing the Riemann solver or the reconstruction method
also affects the dissipation.
The piece-wise linear method to reconstruct the interface states for the
Riemann solver drives in general to higher numerical dissipation because of
the additional interpolation.
In ZEUS the velocity and magnetic fields are already at the interface and the fluxes
can be calculated directly. 
For the inter-code comparison we use the setup described in section 3.1 with a vertical magnetic field with
$n=4$.
Then the resolution in Z is 16 grid cells per
fastest growing mode which is also enough for very diffuse solvers.
For better discerning between the code configurations we include also a weaker vertical 
magnetic field with $n=8$.
\subsection{Measurement method }
To reach the ideal limit and to explain our measurement method we set up a
model with 1 critical mode (64 Grid cells per mode) to reduce the numerical
dissipation (Fig. 3). In addition we use here a critical mode initial disturbance to
enforce the MRI fastest mode to grow faster then any other waves.
Fig. 3 shows the amplitude of the critical mode $\lambda_z$ visible in the radial magnetic field $B_r(Z)$. 
For each radial slice we use the local
orbital time $T_r=2\pi R^{1.5}$ and calculate the growth rate between 1 and 1.5
local orbits.
Most radial slices plotted in Fig. 3 show a growth rate close to the analytical
limit. 
The inner annuli reach non-linear amplitudes first, producing disturbances that
propagate outward and affect the development of the slower-growing
linear modes in annuli further from the star.  As a result the
annuli near R=3 grow faster than indicated by the linear analysis
in all the calculations with sufficient spatial resolution.
%%%%%%%%%%%%%%%%%%%%%%%%%%%%%%%%%%%%%%%%%%%%%%%%%%%%%%%%%%%%%%%%%%%%%%%%%%%%
\section{Inter code comparison}
\subsection{Energy evolution}
\begin{figure}
\hspace{-0.6cm}
\begin{minipage}{5cm}
\psfig{figure=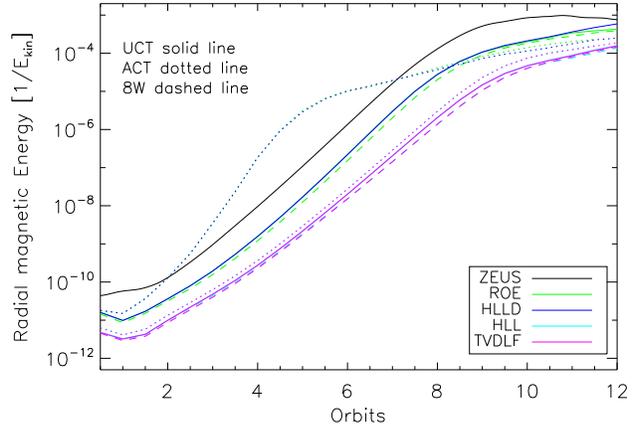,scale=0.50}
\psfig{figure=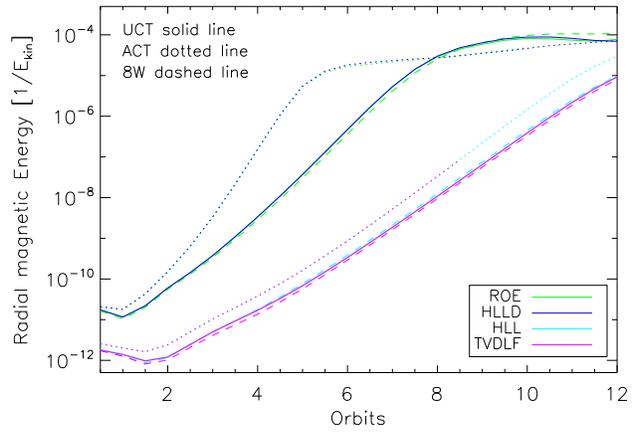,scale=0.50}
\end{minipage}
\label{en4m}
\caption{The total domain integrated radial magnetic energy plotted for n=4 (top) and
n=8 (bottom)
with upwind CT (solid line), Powell 8 wave (dashed line) and arithmetic CT (dotted
line). The values are normalized over the initial orbital kinetic energy. 
After the initial disturbance, the growth of critical mode dominates and there is a linear MRI region
visible between 2 and 8 inner orbits.
For HLLD and Roe, the arithmetic CT shows a growth stronger then MRI until 5 inner orbits caused by a numerical instability.
HLL and TVDLF have a much lower growth because both methods are viscous and do
not resolve the critical mode for the adopted resolution.}
\end{figure}
Table 1 demonstrates the collection of models we made for inter-code comparison.
The linear regime of MRI is visible in the total radial magnetic energy
evolution, Fig. 4. The energy growth due to the excitation
of the critical mode dominates after 2 inner orbits and 
the linear growth regime becomes visible. Fig. 4 shows the radial magnetic energy evolution for the 4 and 8 mode runs. In the 8 mode setups we can
clearly distinguish
3 different classes of evolution. The Lax-Friedrich and HLL Riemann solver with
all possible method of EMF reconstruction (ACT or CT with arithmetic EMF averaging, upwind CT and "8 Wave") have the
slowest energy growth. The second group is built with the HLLD and Roe solvers with the upwind CT and the 8
wave method. Their results are most similar to ZEUS.
The third group are the Roe and HLLD solver in combination with ACT (red
colors in Table 1). The steep slope indicates a numerical problem.
Table \ref{tableoverview4m} presents the growth rate values for the different code configuration.
The growth rates obtained from the model using the ACT scheme are higher
than the analytical value.
\begin{table*}[th]
\begin{center}
\begin{tabular}{l||ccc|ccr}
\hline
&&&&&&\\[-1.5ex]
 Solver  & $ACT_{n=4}$ & $UCT_{n=4}$  & $8W_{n=4}$ & $ACT_{n=8}$ & $UCT_{n=8}$  &
$8W_{n=8}$\\
&&&&&&\\[-1.5ex]
\hline
&&&&&&\\[-1.5ex]
ZEUS &  $0.69 \pm 0.03$ & - & - & $0.69 \pm 0.03$ & - & - \\
HLLD &  \textcolor{red}{$ 0.68 \pm 0.17$}  & $ 0.68 \pm 0.03$ &  -  & \textcolor{red}{$0.86 \pm 0.18$} & $0.70 \pm 0.03$ & -  \\
ROE &  \textcolor{red}{$0.70 \pm 0.18$}  &  $ 0.68 \pm 0.03$ &  $ 0.68 \pm 0.02$ &  \textcolor{red}{$0.84 \pm 0.18$} & $0.70 \pm 0.03$ & $0.70 \pm 0.04$\\
HLL & $ 0.64 \pm 0.05$  &  $ 0.63 \pm 0.05$  &   $ 0.61 \pm 0.05$  &  $0.55 \pm 0.09$ & $0.51 \pm 0.06$ & $0.50 \pm 0.06$\\
TVDLF &  $ 0.64 \pm 0.05$  &  $ 0.63 \pm 0.05$  &  $ 0.63 \pm 0.05$    &
$0.55 \pm 0.09$   & $0.51 \pm 0.06$ & $0.50 \pm 0.06$ \\
%&&&&&&\\[-1.5ex]
\hline
\end{tabular}
\caption{ MRI growth rates are measured for models with 4 and 8 MRI modes
  ($n=4,8$). Measurement of growth rates follows the description in section
  3.3, $\gamma$ is averaged from the different radial shells. 
 The growth rates higher than the analytical value are marked with red. The uncertainty numbers
are measured from the radial variations in the growth rate visible in Fig. 7 or Fig.
9.}
\label{tableoverview4m}
\end{center}
\end{table*}

\subsection{Arithmetic CT}
In case of ACT and the high-resolution solver, the strong growth of magnetic energy
overshoots the analytical expectation for MRI growth (Table 1 and Fig. 4). 
Here a numerical instability dominate s in the cases with the HLLD and Roe solvers.
The fastest-growing mode has a vertical wavelength shorter than in
the linear analysis.
In Fig. 5 we plotted the growth rate for each radial slice, calculated from 1
to 1.5 local orbits. Roe and HLLD show the same unphysical high growth rate.
On first inspection, the more diffusive
HLL and TVDLF solvers seem not to be affected.  However, at higher
resolutions the numerical instability occurs there as well.
The instability shows a 'checkerboard' pattern (Fig. 5).
The occurrence of such instabilities was already mentioned in Miyoshi \& Kusano (2008).
The Roe and HLLD solvers which include the Alfv\'en
characteristic accurately evolve the disturbances independently of
the resolution.
The reason of the numerical instability lies in the inconsistent EMF reconstruction,
as described in \citet{lon00}.
The correction methods are proposed in \citet{lon04}, \citet{gar05} and \citet{zie04}.
In short one can summarize, that the simple arithmetic average of 4 Godunov fluxes to reconstruct the EMF is not consistent with the upwind scheme.
The error becomes visible in case of dominating flows along the azimuthal direction in
3D, presented in the Loop advection test from \citet{gar05}.
It is interesting to note that the results of HLLD and Roe are very close. Also the values of HLL and TVDLF are 
exact the same in our tests. Resolving the Alfv\'en characteristic in MHD
Riemann solver plays an important role
for numerical dissipation.
\begin{figure}
\hspace{-0.6cm}
\begin{minipage}{5cm}
\psfig{figure=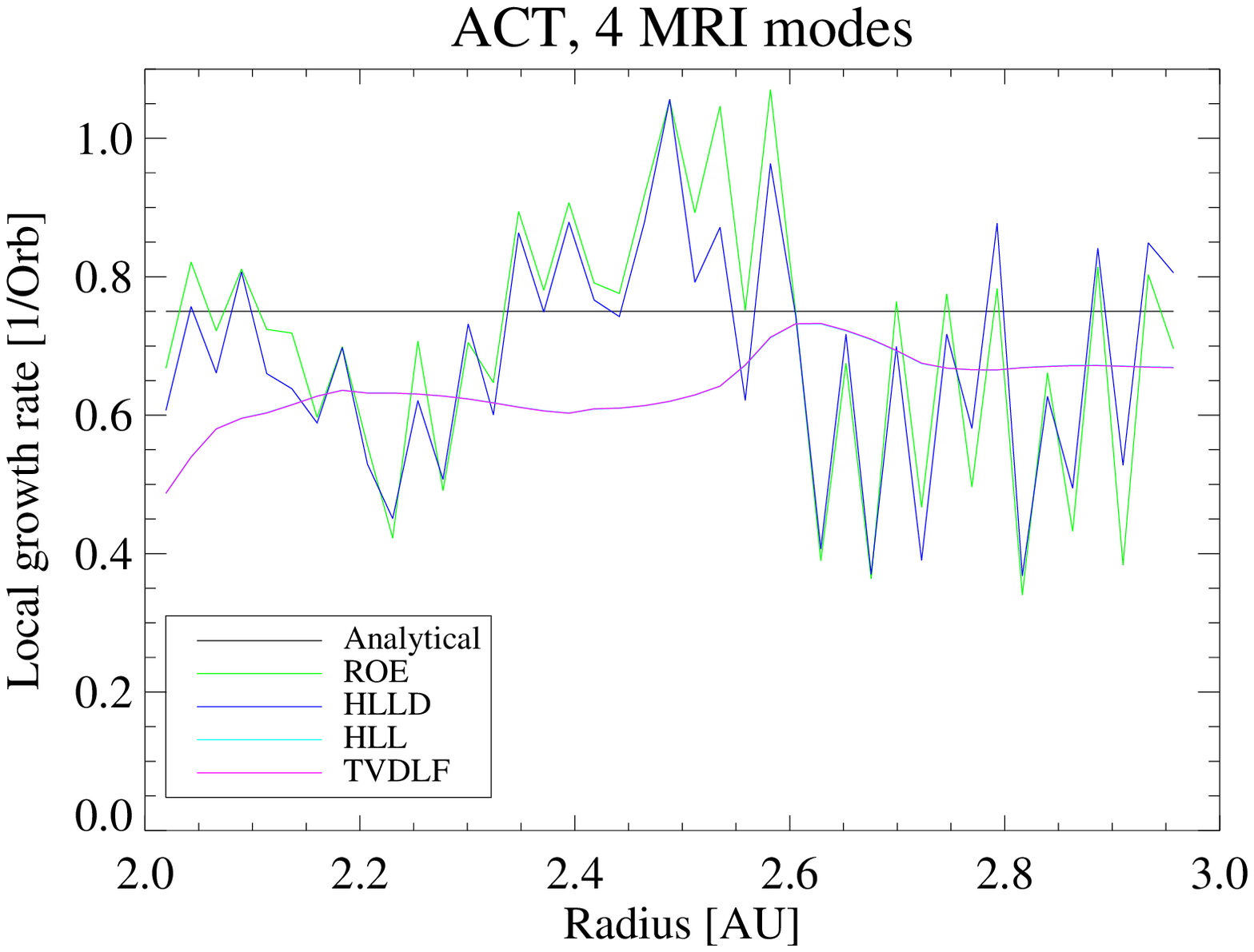,scale=0.50}
\psfig{figure=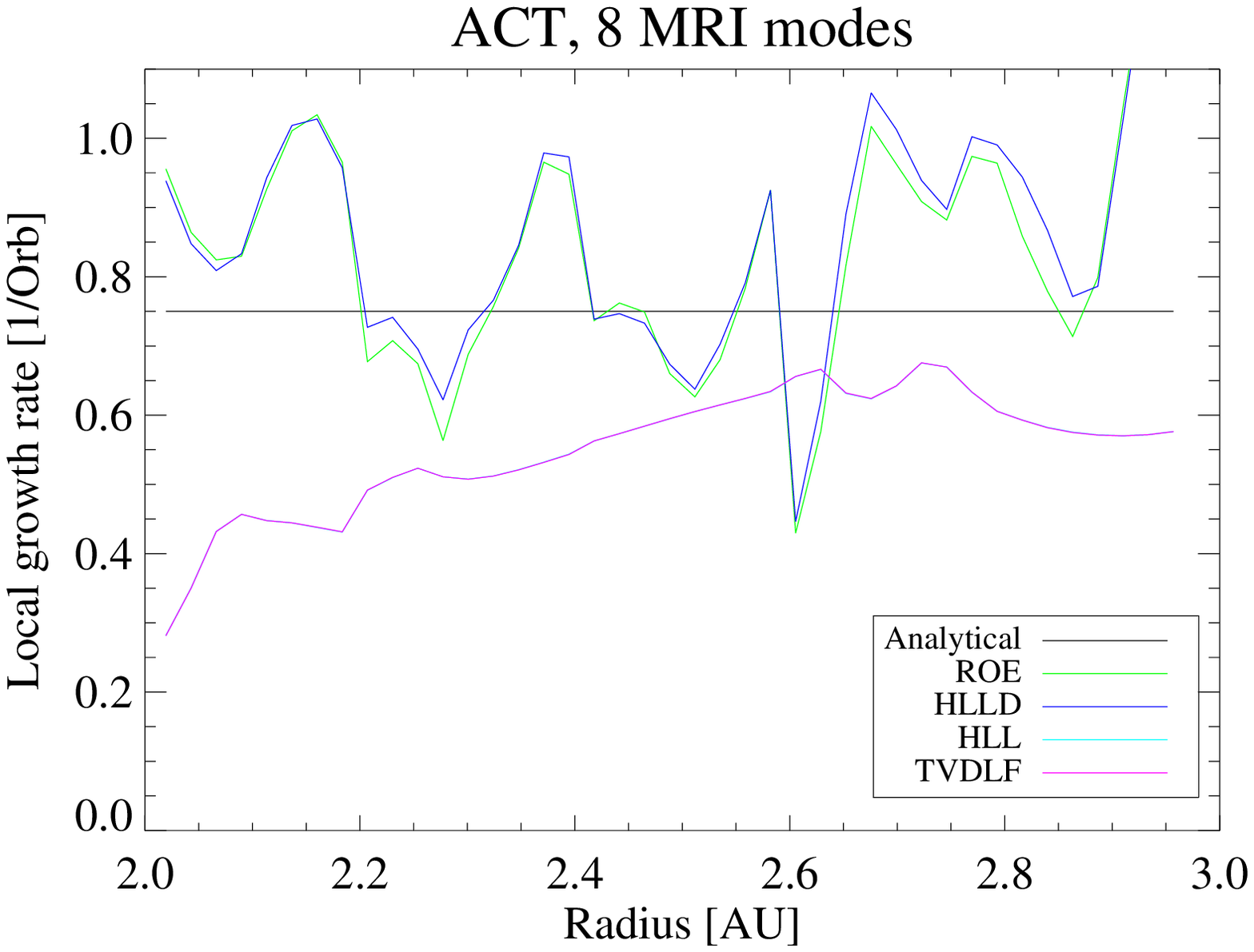,scale=0.50}
\end{minipage}
\label{act4m}
\caption{Local growth rate for n=4 (top) and n=8 (bottom) with arithmetic CT. For HLLD and Roe the local growth rate exceeds the analytical limit. The problem does not affect the HLL and TVDLF solver for the used resolution} %. They have nearly exact evolution.}
\end{figure}
\begin{figure}
\hspace{-0.6cm}
\begin{minipage}{5cm}
\psfig{figure=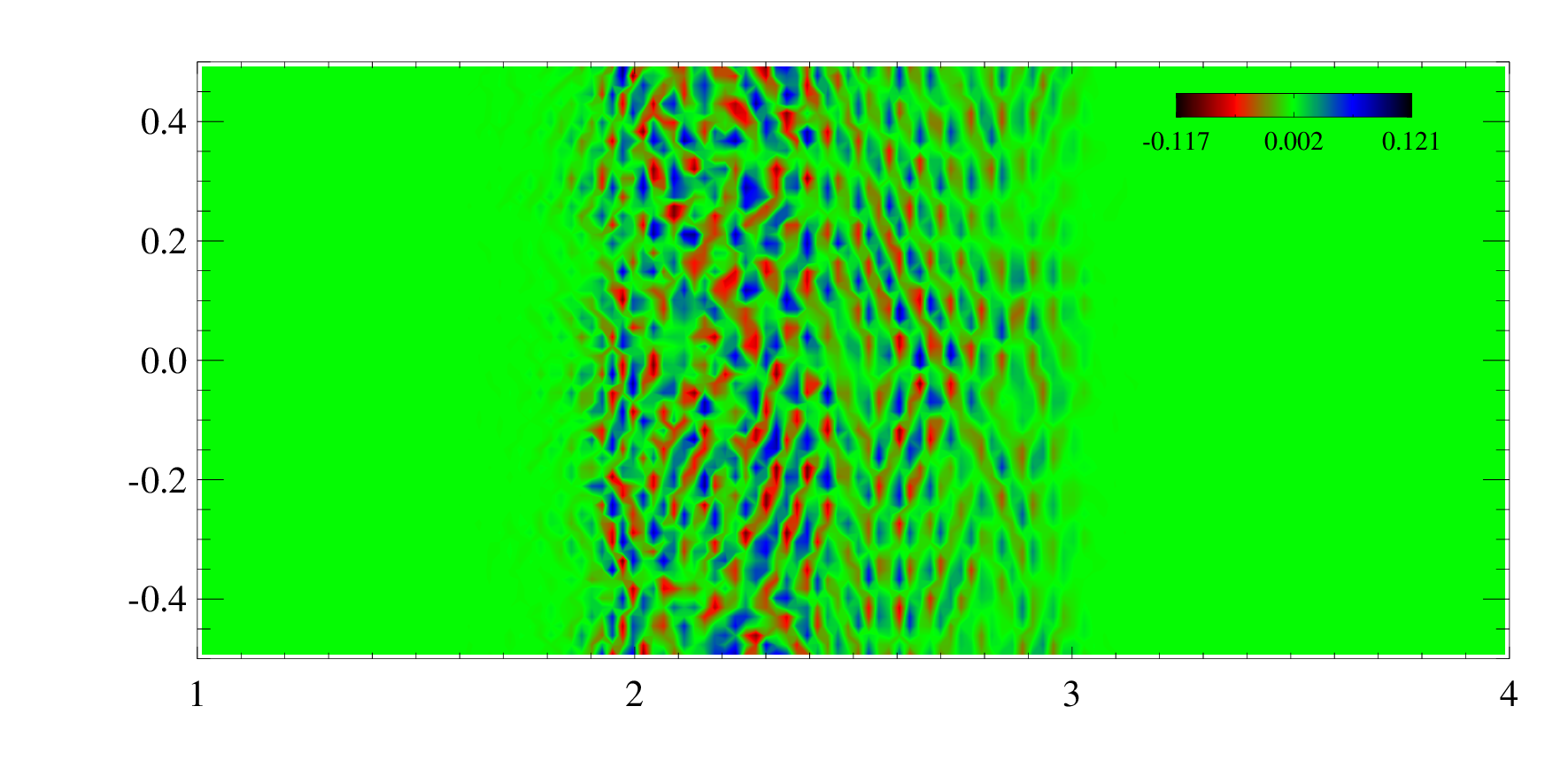,scale=0.32}
\psfig{figure=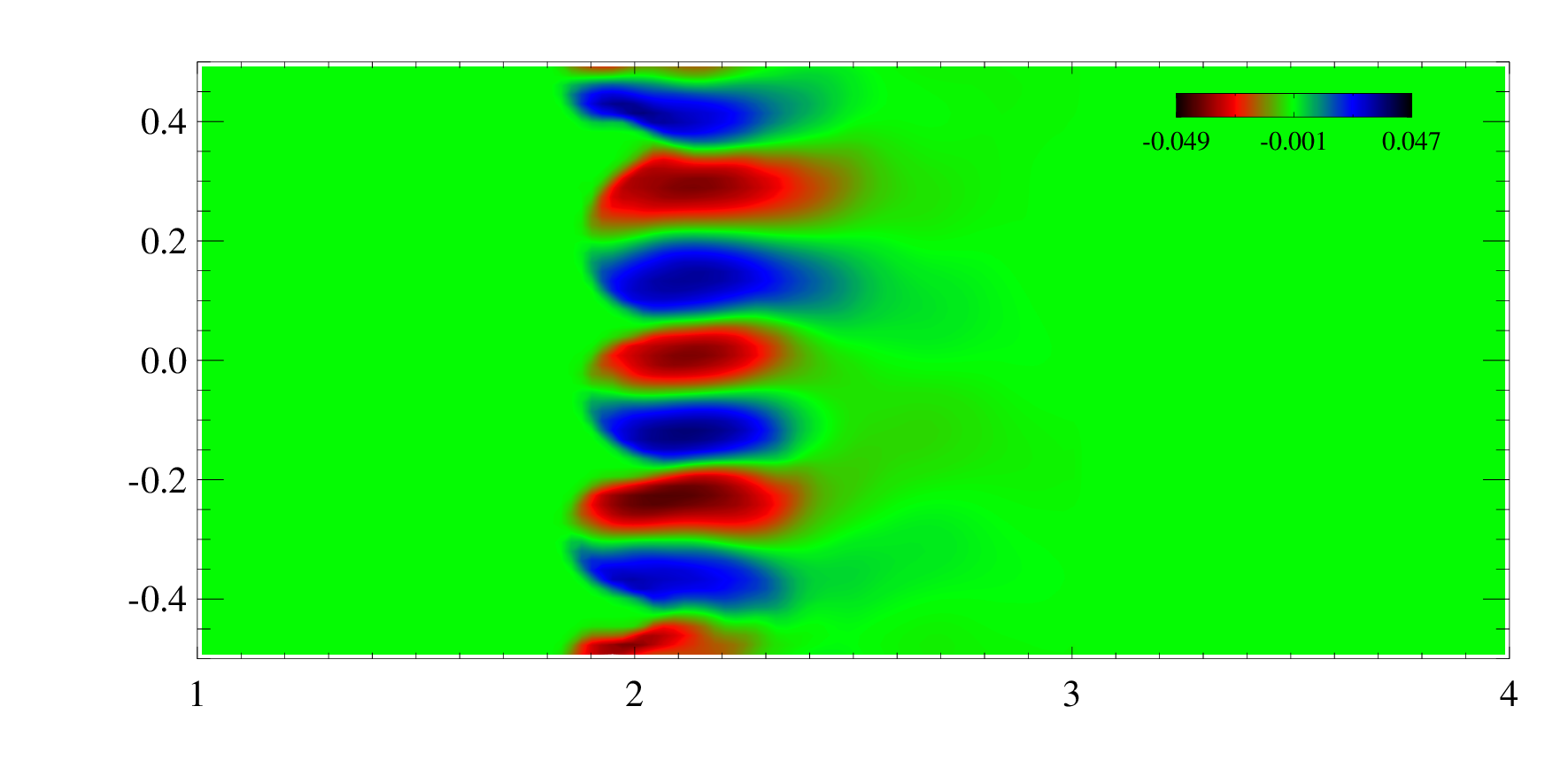,scale=0.32}
\psfig{figure=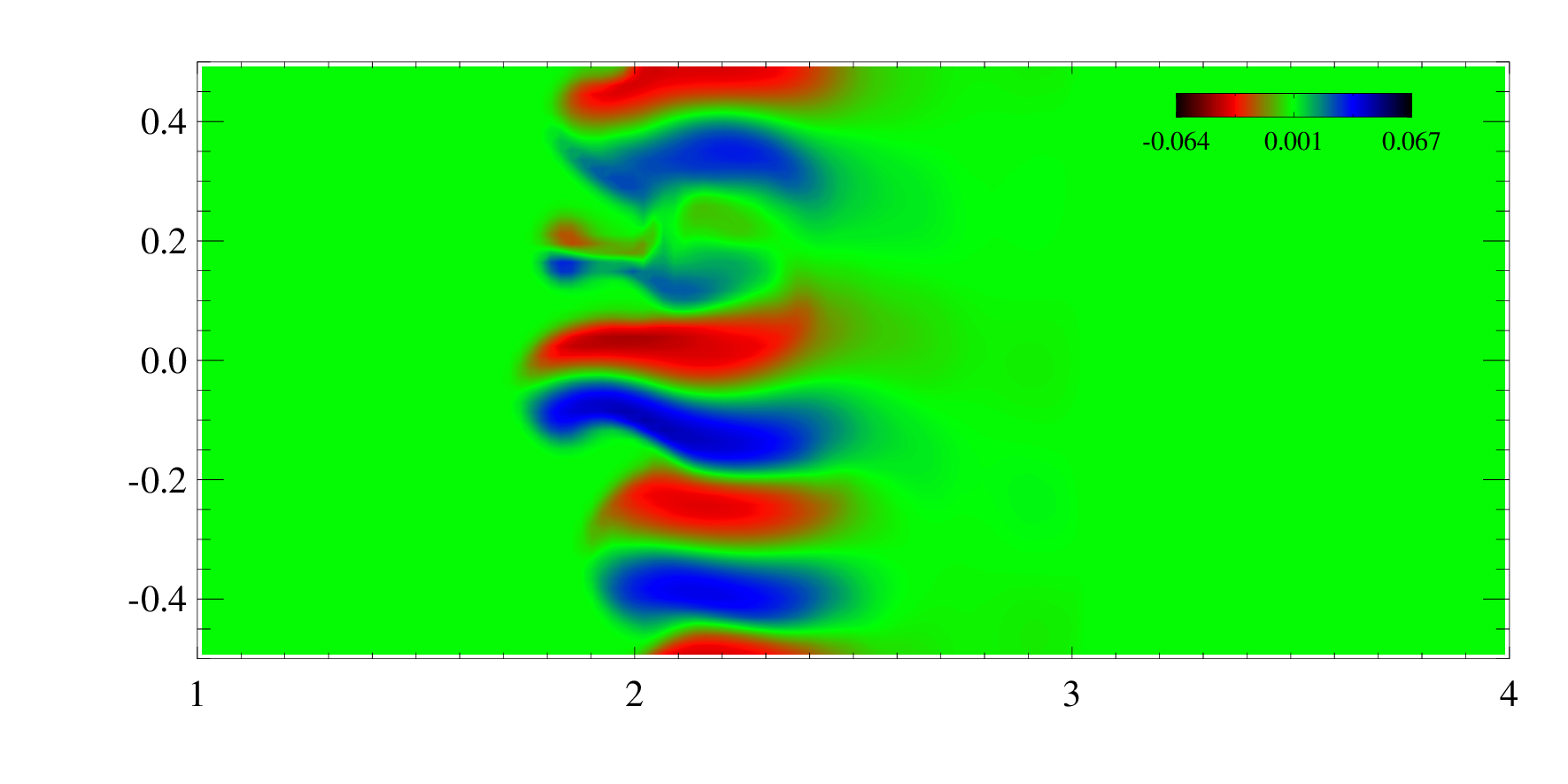,scale=0.32}
\end{minipage}
\label{4mcont}
\caption{Contour plot of azimuthal magnetic field for arithmetic CT (top), UCT (middle) and ZEUS (bottom). In case of arithmetic CT, the numerical instability
dominates MRI. In case for UCT and ZEUS the n=4 is visible. A change from blue to red means a change of
magnetic field sign. The snap-shot of magnetic field is taken after 8 inner orbits.}
\end{figure}
%(#### PLOT INSTABILITY  ####)
\subsection{Upwind CT vs 8 wave}
With the Upwind CT EMF reconstruction the critical MRI mode dominates 
 and 'checkerboard' pattern disappears.
Fig. 6 presents a contour plot of the azimuthal magnetic field. 
HLLD with upwind CT and ZEUS show 4 evolving modes. 
Fig. 7 shows the small difference between "8 Waves" and upwind CT.
The additional EMF corner-integration $\rm{UCT}_{CONTACT}$ does not change the result.
The strongest effect on numerical dissipation has the change of Riemann solver. 
\begin{figure}
\hspace{-0.6cm}
\begin{minipage}{5cm}
\psfig{figure=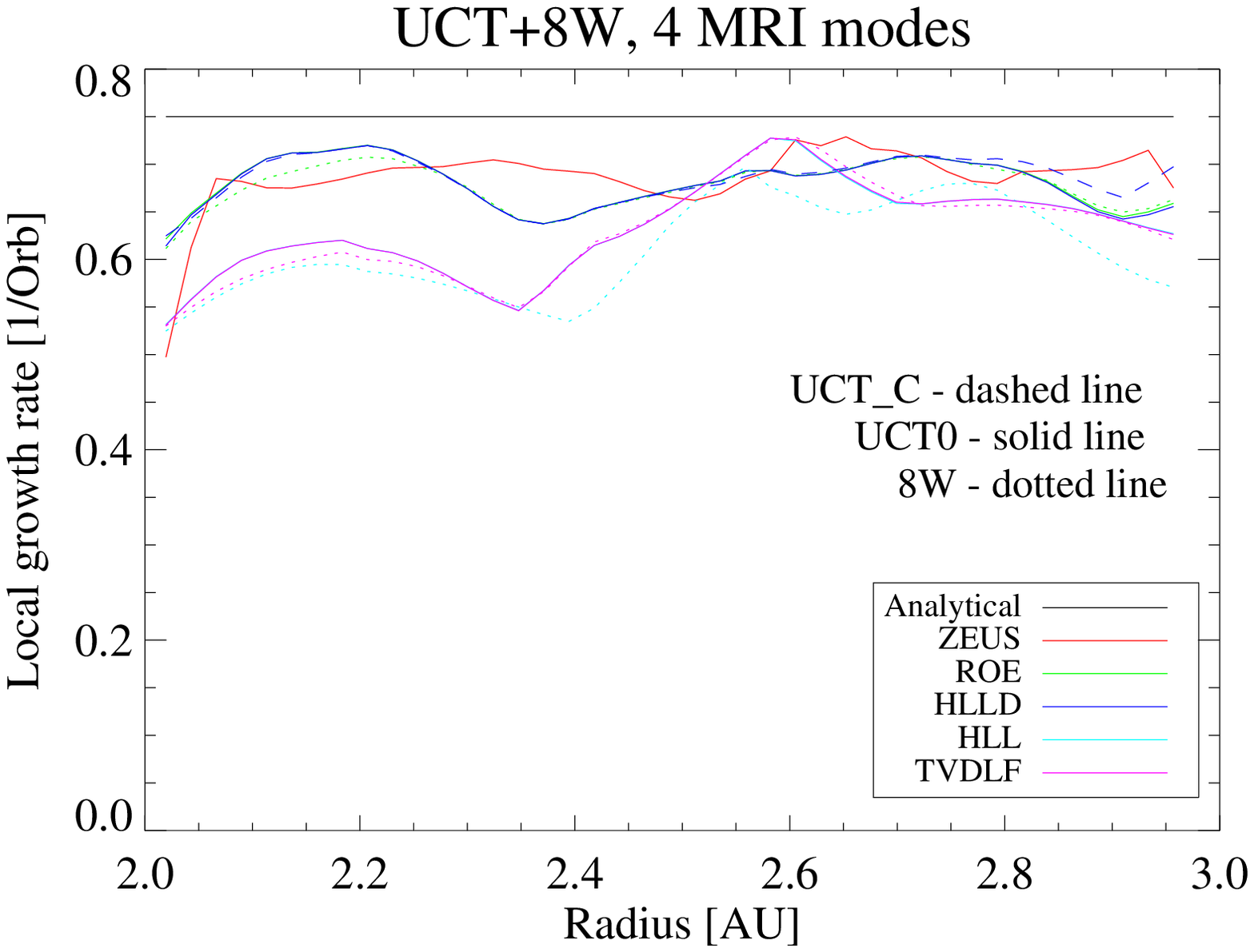,scale=0.50}
\psfig{figure=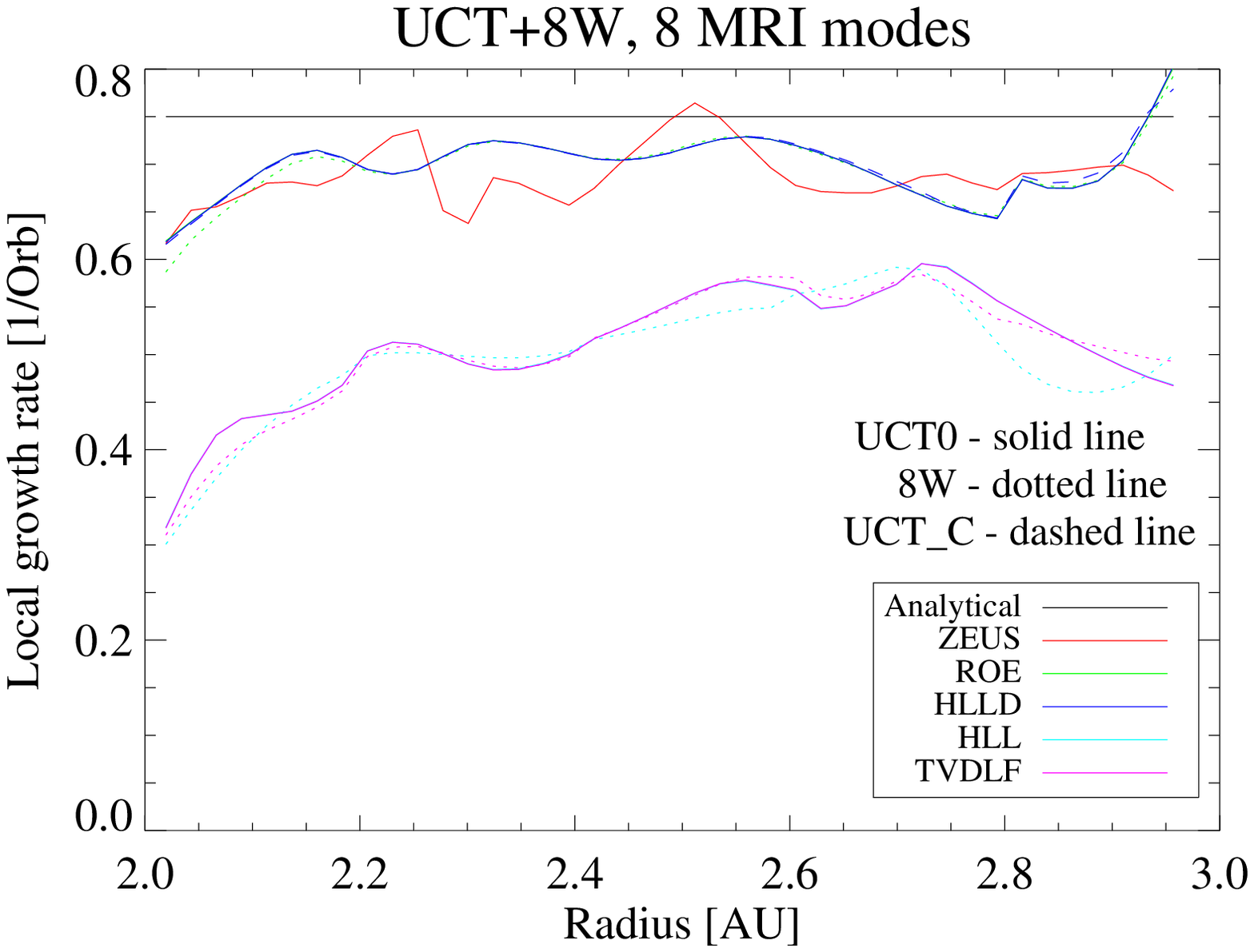,scale=0.50}
\end{minipage}
\label{uct4m}
\caption{Local growth rate for n=4 (top) and n=8 (bottom) with upwind CT (solid line), $\rm{UCT}_{\rm{contact}}$ (dashed line) and Powell's Eight Wave (dotted line). The high resolution MHD Riemann Solver HLLD and Roe show the same MRI evolution, producing same growth rate as ZEUS. HLL and TVDLF are more diffusive.}
\end{figure}
The reason is the number of wave characteristics which is used by each solver.
HLL and TVDLF use the fast magneto-sonic wave characteristic as maximal signal velocity. 
HLLD includes additional the Alfv\'en wave characteristic and the contact discontinuity. The ROE solver includes all 7 MHD waves,
as well as the slow magneto-sonic characteristic.
ZEUS shows the MRI development same as HLLD and Roe.
\subsection{4 to 8 MRI modes }
%As mention in section 3.2, 
In case of 4 modes, the resolution in the
Z direction is sufficient for all Riemann
solvers to resolve the critical mode.
HLLD , Roe and ZEUS shows the highest growth rate followed by TVDLF and HLL.
As already presented in Fig. 4. 
The models for 8 MRI critical modes are summarized in Table 1. Energy evolution is shown in Fig.4 (bottom).
Growth rates are significantly reduced only for HLL and TVDLF solvers.
ZEUS and the MHD Riemann solver ROE and HLLD resolve 7 modes, HLL and TVDLF only 4 modes.
The resolution about 8 grid cells per critical mode is not enough to resolve the modes perfectly.
Nevertheless, the additional Alfv\'en characteristic included in the HLLD solver drives to lower numerical diffusion.
Fig. 8 shows a contour plot of radial magnetic field after 7 inner orbits for HLLD solver and ZEUS.
The ZEUS code shows the MRI pattern very close to those of HLLD and Roe.
\begin{figure}
\hspace{-0.6cm}
\begin{minipage}{5cm}
\psfig{figure=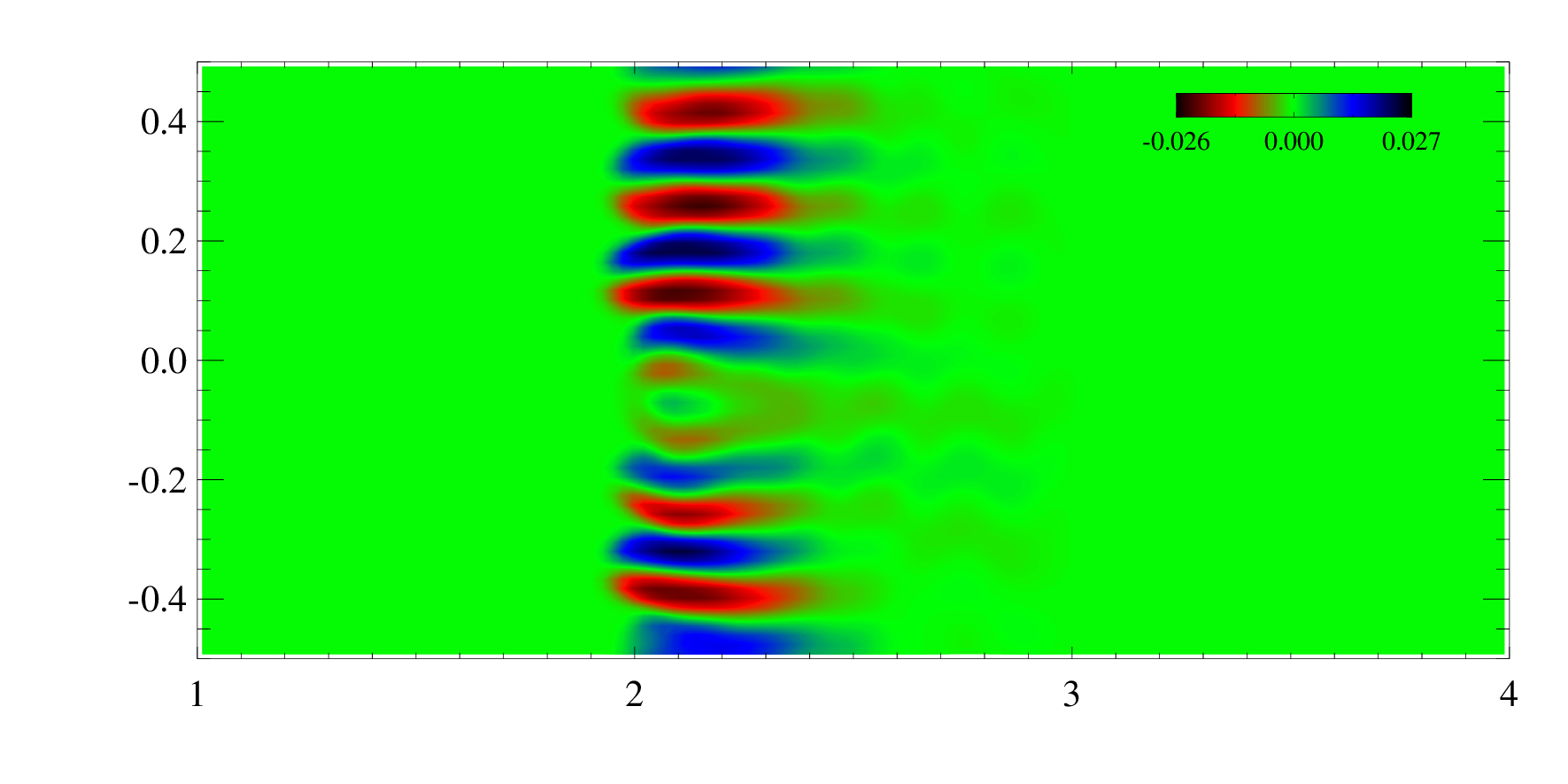,scale=0.32}
\hspace{0.5cm}
\psfig{figure=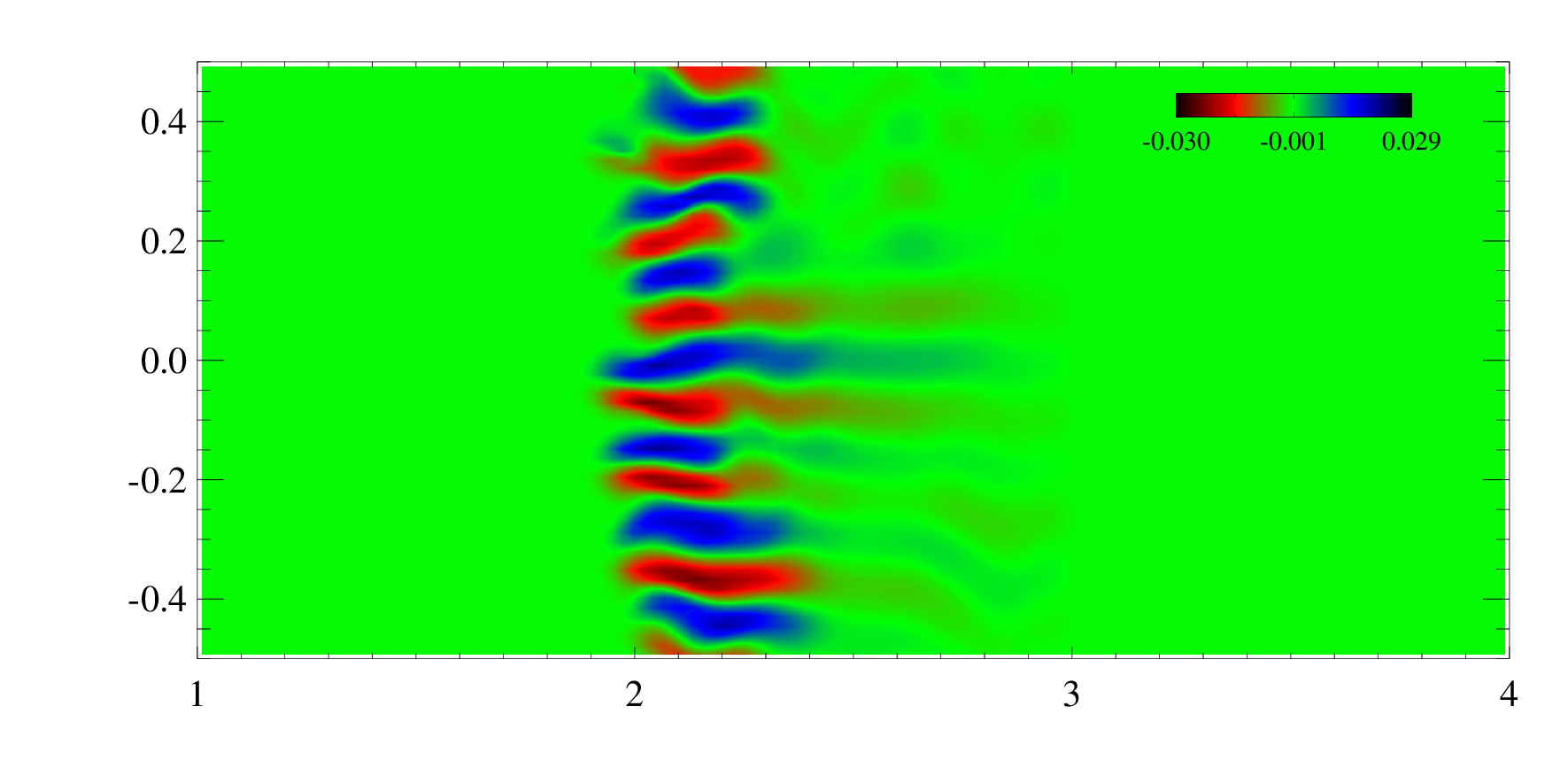,scale=0.32}
\end{minipage}
\label{brad8mode}
\caption{Contour plot of radial magnetic field in code units for the HLLD solver (top) and ZEUS (Bottom). After 7 inner
orbits the inner radial part breaks into nonlinear regime. This affects also the growth rate at the outer part
as we have already seen.}
\end{figure}

\subsection{PPM vs PLM}
The reconstruction of the states at the interfaces
leads to the high dissipation in the Godunov schemes (see section 2). 
Therefore we test the fourth order piece-wise parabolic method (PPM) by Collela (1984) 
for the ROE,HLLD and for the diffusive TVDLF solver.
\begin{figure}
\hspace{-0.6cm}
\begin{minipage}{5cm}
\psfig{figure=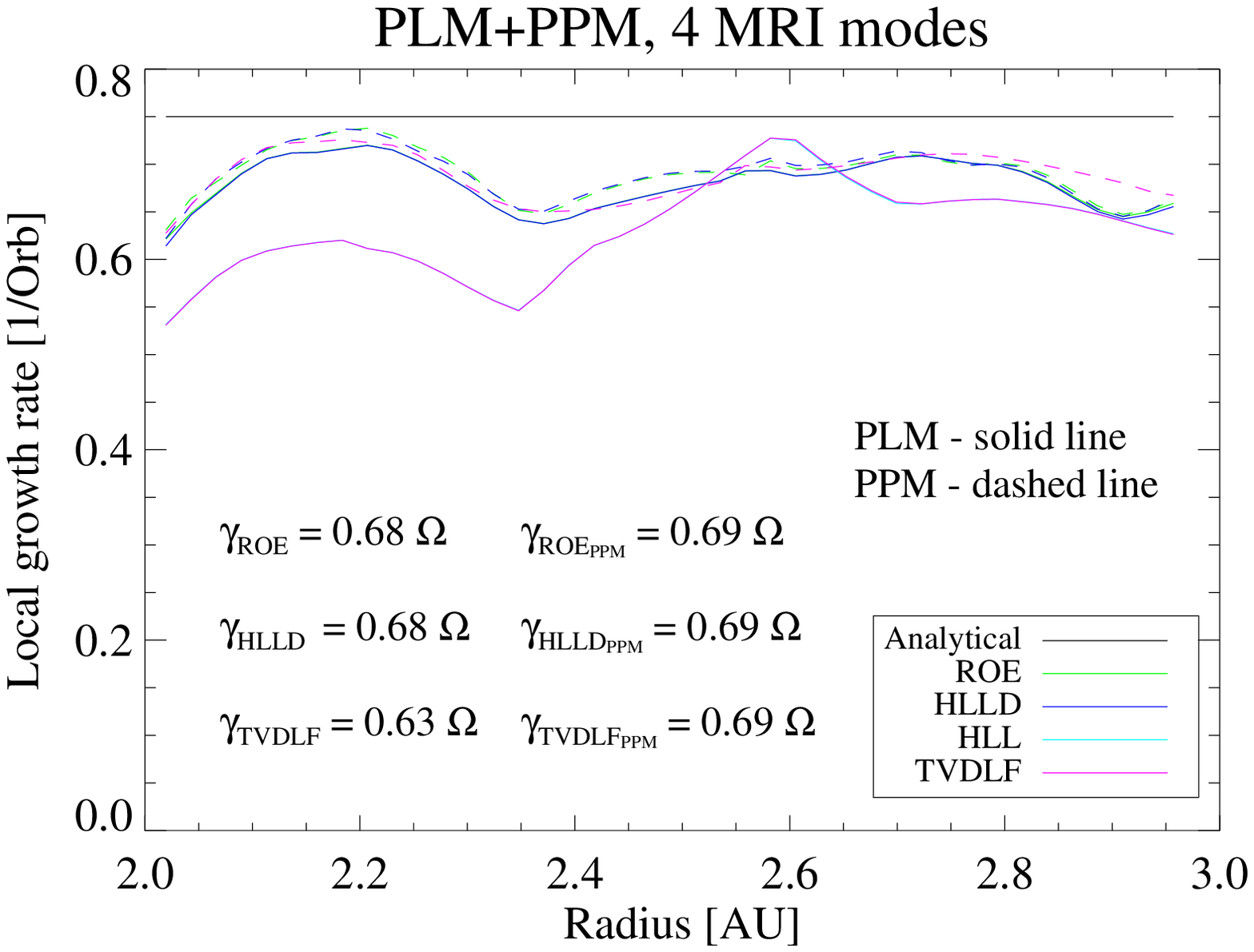,scale=0.50}
\psfig{figure=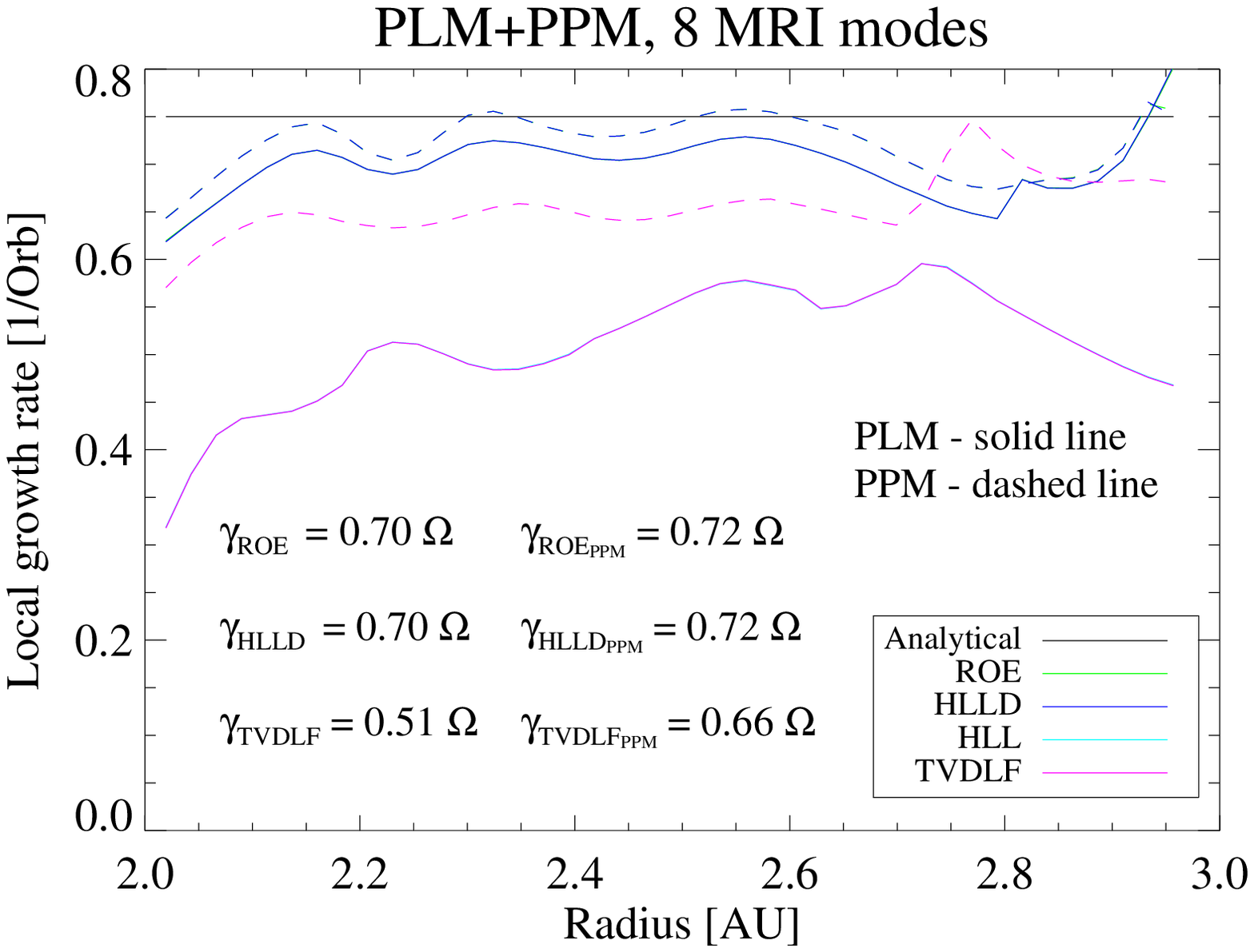,scale=0.50}
\end{minipage}
\label{ppmvslin}
\caption{MRI growth rate for PPM (dashed) and PLM (solid). 
In the case of the Lax-Friedrich solver, the PPM
interpolation leads to a lower numerical dissipation and growth
rates nearer the analytic solution.
Differences between HLLD and ROE solvers are negligible.}
\end{figure}
Fig. 9 shows the difference in local MRI growth rate for simulations made for interpolations with piece-wise linear method (PLM) and PPM. The results of MHD Riemann solver HLLD and ROE are not
strongly affected, though 
there is a slightly lower numerical dissipation visible when PPM is used. But there is a strong effect of the interpolation method visible for the classical Lax-Friedrich and HLL solvers. 
In case of the 8 mode run,  also the classical Riemann solver presents 7 modes and the
values of ROE, HLLD and TVDLF converge. 
To verify this conclusion we investigate the resolution convergence, including
only the HLLD and TVDLF solver with upwind CT and both reconstruction
methods.
\section{Convergence test}
The models for the convergence tests are summarized in Table 2. We take here
the n=4 model.
We double the resolution up to 4 times for each dimension to check for the earliest convergence.
In these runs we use an initial axisymmetric white noise for the radial and vertical
velocity (\textit{www.random.org}).
This leads to slightly different growth rate values as in Table 1.
Fig. 10 shows the results. The HLLD MHD Riemann solver in combination with
piece-wise parabolic reconstruction 
shows the earliest convergence.
Even with a resolution about 32 in Z and with the PPM reconstruction, we can resolve the critical mode
and get the converged value of about $\gamma=0.70\Omega$ (Table 2). The resolution runs marked with a star do
not resolve the fastest growing MRI mode.
As expected, the Lax Friedrich solver shows the highest numerical dissipation.
Here 8 grid cells per fastest growing mode are not enough.
For the highest resolution of
512x256x256 we see in all cases a decreasing of the growth rate. 
We cannot say whether
this behavior is connected with the presence of high radial
wavenumber or a change in the numerical dissipation.  All except
the Zeus runs decline by less than the uncertainty.
\begin{table*}[th]
\begin{tabular}{l|ccccr}
\hline
&&&&&\\[-1.5ex]
 Solver  &      32x16x16 & 64x32x32 & 128x64x64 & 256x128x128 & 512x256x256 \\
&&&&&\\[-1.5ex]
\hline
\hline
$\rm{ZEUS}_{\rm{LIN}}$ &   0.58*$\pm$0.07 & 0.65$\pm$0.05  & 0.70$\pm$0.07  & 0.68$\pm$0.05& 0.62$\pm$0.05 \\
$\rm{HLLD}_{\rm{LIN}}$ &    0.53*$\pm$0.07 & 0.65$\pm$0.07 & 0.70$\pm$0.05 & 0.70$\pm$0.03 & 0.68$\pm$0.05\\
$\rm{HLLD}_{\rm{PPM}}$ &  0.60*$\pm$0.05 & 0.70 $\pm$0.06 & 0.71$\pm$0.05 & 0.71$\pm$0.04 & 0.67$\pm$0.04\\
$\rm{TVDLF}_{\rm{LIN}}$ &   0.33*$\pm$0.05 & 0.53*$\pm$0.09 & 0.65$\pm$0.05 & 0.70$\pm$0.04 & 0.67$\pm$0.04\\
$\rm{TVDLF}_{\rm{PPM}}$ &  0.50*$\pm$0.05 & 0.63*$\pm$0.08 & 0.72$\pm$0.06 & 0.70$\pm$0.03 & 0.67$\pm$0.04\\ 
\hline 
\end{tabular}
\caption{ Growth rates for the resolution tests. The asterisks mark the runs
which was not able to resolve n=4.}
\label{conv}
\end{table*}
\begin{figure}
\hspace{-0.6cm}
\begin{minipage}{5cm}
\psfig{figure=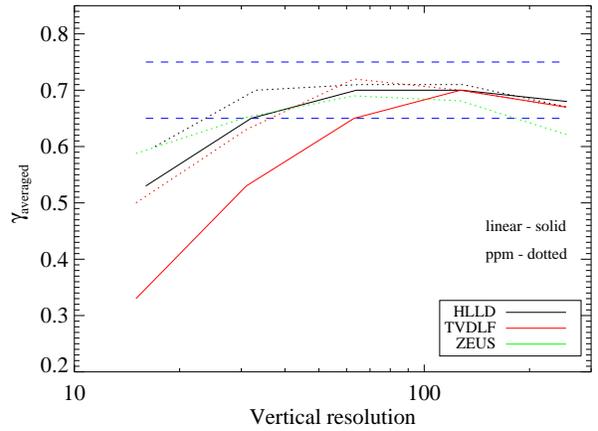,scale=0.50}
\end{minipage}
\label{convfig}
\caption{
Growth rates exhibit a quick convergence to the $\gamma \sim 0.7$ 
with increasing resolution for all Godunov solvers, with the
TVDLF solver performing notably worse.
 The dashed lines are marking the uncertainty of growth rate measurements. 
A decline is apparent at the highest resolution.
}
\end{figure}
\newpage
%%%%%%%%%%%%%%%%%%%%%%%%%%%%%%%%%%%%%%%%%%%%%%%%%%%%%%%%%%%%%%%%%%%%%%%%%
\section{Conclusion}
%%%%%%%%%%%%%%%%%%%%%%%%%%%%%%%%%%%%%%%%%%%%%%%%%%%%%%%%%%%%%%%%%%%%%%%%%%%%
The inter-code comparison and the convergence test stress the importance
of treating the Alfv\'en characteristics for MHD Riemann solver in MRI simulations.
Observing the consequences implied by numerical dissipation, we
obtain a robust and accurate Godunov scheme for 3D MHD simulation in curvilinear coordinate systems. 
The arithmetic average of the 4 Godunov fluxes \citep{bal99} is not consistent 
with the upwind Godunov scheme \citep{lon00} and leads to numerical instability.
Beside the loop advection test \citep{gar08}, the linear MRI evolution in global disks is a crucial test for a consistent MHD method.
Our 3D MHD simulation with ACT show the unphysical behavior of a
'chessboard' instability, mentioned for the HLLD solver in \citet{miy08}.
The high-order MHD Riemann solvers HLLD and ROE, which contain the Alfv\'en characteristic, 
reproduce the 'chessboard' pattern in the classical ACT scheme independent of resolution. 

The upwind EMF reconstruction in combination with the accurate and robust HLLD MHD Riemann solver
represents a good tool for future simulations of 3D MHD accretion disks.
We reproduce the linear MRI analysis in the inter-code comparison.
The Lax-Friedrich and HLL solvers demonstrate much higher numerical dissipation in comparison to the finite difference scheme ZEUS.
The high-order MHD Riemann solver HLLD and ROE can compensate this dissipation by
 using more MHD characteristics. The Alfv\'en characteristic appears to play an important role for MRI.
  For the linear MRI evolution both HLLD and ROE show the same
numerical dissipation as the finite difference scheme ZEUS. In addition, the HLLD solver
shows nearly the exact evolution like ROE, even without resolving the slow magneto-sonic characteristic.
In combination with the piece-wise parabolic method the HLLD solver has the earliest convergence compared to the Lax-Friedrich solver and ZEUS.
\section{Outlook}
The Prandtl numbers in global
simulations for the high-order MHD Riemann solver and physical dissipation
 are very important for MRI turbulence in proto-planetary disks.
Proper MHD tests shall be done, exploring the numerical dissipation of the PLUTO code. 
\citet{froI07} measured the numerical Prandtl number for 
ZEUS, which is 
 in the region from 2 to 8.
Recent work by \citet{sim09} presented a Prandtl number about 2 for
the Godunov type ATHENA code. 
Nevertheless, in the proto-planetary disks one can expect rather smaller Prandtl numbers, $\rm Pm<1$.
The consequences of numerical $\rm Pm$ for MRI shall be explored.
\paragraph{\textit{Acknowledgments.}}
%\begin{acknowledgments}
We thank Udo Ziegler for the very useful discussions about the upwind Constrained transport property.
We also thank Neal Turner, Thomas Henning and Matthias Griessinger for the post process of this work.
H. Klahr, N. Dzyurkevich and M. Flock have been supported in part by the
Deutsche Forschungsgemeinschaft DFG through grant DFG Forschergruppe 759
”The Formation of Planets: The Critical First Growth Phase”. Parallel
computations have been performed on the PIA cluster of the Max-Planck
Institute
for Astronomy Heidelberg located at the computing center of the Max-Planck
Society in Garching.
%\end{acknowledgments}

%%%%%%%%%%%%%%%%%%%%%%%%%%%%%%%%%%%%%%%
%\bibliographystyle{aa} % style aa.bst
%\bibliography{GMRI} % your references Yourfile.bib

%---------------------------------------------------------------
%---------------------------------------------------------------

\end{document}